\documentclass[a4paper,fleqn,usenatbib]{mnras}

\usepackage{newtxtext,newtxmath}

\usepackage[T1]{fontenc}
\usepackage{ae,aecompl}

\usepackage{graphicx,color}	
\usepackage{amsmath}	

\def\msun{M$_{\odot}$}
\def\msunyr{M$_{\odot}$\,yr$^{-1}$}

\definecolor{Mygrey}{gray}{0.75}

\newcommand{\ltsimeq}{\raisebox{-0.6ex}{$\,\stackrel{\raisebox{-.2ex}{$\textstyle <$}}{\sim}\,$}}

\mathchardef\mhyphen="2D

\title[IMF variation and Deuterium]{Stellar initial mass function variation in massive early-type galaxies: the potential role of the deuterium abundance}
\author[Timothy A. Davis \& Freeke van de Voort]{
Timothy A. Davis,$^{1}$\thanks{E-mail: DavisT@cardiff.ac.uk}
and Freeke van de Voort$^{1}$
\\
$^{1}$School of Physics \&\ Astronomy, Cardiff University, Queens Buildings, The Parade, Cardiff, CF24 3AA, UK}

\date{Accepted 2020 September 01. Received 2020 August 28; in original form 2020 July 16}

\pubyear{2020}

\begin{document}
\label{firstpage}
\pagerange{\pageref{firstpage}--\pageref{lastpage}}
\maketitle

\begin{abstract}
The observed stellar initial mass function (IMF) appears to vary, becoming bottom-heavy in the centres of the most massive, metal-rich early-type galaxies.  It is still unclear what physical processes might cause this IMF variation. In this paper, we demonstrate that the abundance of deuterium in the birth clouds of forming stars may be important in setting the IMF. We use models of disc accretion onto low-mass protostars to show that those forming from deuterium-poor gas are expected to have zero-age main sequence masses significantly lower than those forming from primordial (high deuterium fraction) material. This deuterium abundance effect depends on stellar mass in our simple models, such that the resulting IMF would become bottom-heavy -- as seen in observations. Stellar mass loss is entirely deuterium-free and is important in fuelling star formation across cosmic time. Using the EAGLE simulation we show that stellar mass loss-induced deuterium variations are strongest in the same regions where IMF variations are observed: at the centres of the most massive, metal-rich, passive galaxies. While our analysis cannot prove that the deuterium abundance is the root cause of the observed IMF variation, it sets the stage for future theoretical and observational attempts to study this possibility.
\end{abstract}




\section{Introduction}

One of the key techniques in an extragalactic astrophysicist's toolbox is the ability to link the observed electromagnetic light from galaxies to their physical quantities (such as mass, star formation rate etc.). A key assumption in this process is the use of a stellar initial mass function (IMF), which is typically assumed to be invariant regardless of environment, and over the entirety of cosmic history. Over the past decade, however, evidence has mounted that the IMF may vary in galaxies outside of our own, both in passive early-type galaxies, and highly star-forming objects. Understanding the magnitude of this variation, and the fundamental physics underlying it, is crucial to allow us to make progress.

Massive early-type galaxies in the local universe have been suggested to contain more mass per unit light than expected assuming a \cite{2001MNRAS.322..231K} or \cite{2003PASP..115..763C} IMF. Through dynamical studies \citep[e.g.][]{2012Natur.484..485C,2013MNRAS.432.2496D,2013ApJ...765....8T,2015MNRAS.446..493P,2017ApJ...838...77L}, gravitational lensing \citep[e.g.][]{2010ApJ...709.1195T,2010ApJ...721L.163A}, and stellar population analysis \citep[e.g.][]{2003MNRAS.339L..12C,2010Natur.468..940V,2012ApJ...760...71C,2012ApJ...753L..32S,2013MNRAS.429L..15F,2013MNRAS.433.3017L,2019A&A...626A.124M,2019MNRAS.485.5256Z}, various authors have shown that this effect is unlikely to be due to dark matter and is typically strongest in the central regions of these systems \citep[e.g.][]{2015MNRAS.447.1033M,2017MNRAS.464..453D,2017ApJ...841...68V,2018MNRAS.474.4169O,2018MNRAS.477.3954P}. Thus it seems that the IMF in these systems must vary. Dynamical analyses (such as modelling of stellar/gas kinematics) and lensing are unable to ascertain if this extra mass is in the form of low-mass stars or stellar remnants. Stellar population analyses, especially those which probe gravity-sensitive spectral indices, suggest that this excess mass is present in low-mass stars \citep[e.g.][]{2010Natur.468..940V}. Although a range of uncertainties can affect all of these measurements, and vital cross-checks between different techniques are still ongoing \citep[e.g.][]{2014MNRAS.443L..69S,2017MNRAS.468.1594A,2017ApJ...845..157N,2018MNRAS.475.1073V,2018MNRAS.478.1595C}, a consensus seems to be emerging that the IMF is \textit{bottom-heavy} in massive ETGs (see \citealt{Smith_ARAA2020} for a recent review). 

Another place where a changing IMF has been invoked is in starburst galaxies. It has been suggested that the most star-forming systems, both in the local universe and at high redshift, require an excess of high-mass stars. Without such an excess various authors are unable to explain the level of emission seen in high-mass star formation tracers, such as the equivalent width of the H$\alpha$ emission line (e.g. \citealt{1983ApJ...272...54K,1994ApJ...435...22K,2008ApJ...675..163H,2011MNRAS.415.1647G,2017MNRAS.468.3071N}), or the H$\alpha$-to-UV ratio (e.g. \citealt{2009ApJ...695..765M,2009ApJ...706..599L,2009ApJ...706.1527B}). Observations of the chemical enrichment of certain isotopes by high-mass star formation also suggests the need for more high-mass stars, and a \textit{top-heavy} IMF in starburst galaxies \citep[e.g.][]{2018Natur.558..260Z,2019ApJ...879...17B}.

The cause of these putative variations in the IMF is not clear. Various observations have been used to show that the IMF becomes progressively more bottom-heavy in higher velocity dispersion \citep[e.g.][]{2012Natur.484..485C,2013MNRAS.433.3017L,2015MNRAS.446..493P}, higher metallicity \citep[e.g.][]{2015ApJ...806L..31M,2018MNRAS.477.3954P,2019MNRAS.485.5256Z}, and more alpha-element enriched \citep[e.g.][]{2012ApJ...760...71C} early-type galaxies. This has been used to suggest that the altered IMF we observe today was set at high redshift, in the extreme bursts of star formation that formed the progenitors of today's massive ellipticals. This conclusion is seemingly in conflict with the observations suggesting a top-heavy IMF in such strongly star-forming systems.  Gas density, metallicity, temperature, and velocity dispersion can all affect the jeans length and fragmentation within star-forming clouds. The right combination of these parameters have been shown in theoretical works to create both top-heavy and bottom-heavy IMFs \citep[e.g.][]{2005MNRAS.356.1201B,2007ApJS..169..239G,2013ApJ...770..150H}. It is not clear, however, whether an IMF that scales with any of these cloud scale properties can match all the available observations \citep[e.g.][]{2017ApJ...845..136B,2019MNRAS.485.4852G}. Given the above inferences and contradictions, as well as the range of uncertainties that affect the measurements, it is unclear which (if any) physical parameter so far considered actually induces the variation in the IMF.  

In this paper we aim to demonstrate that a further parameter may be important in setting the IMF: the abundance of deuterium in the birth clouds of forming stars. {To the best of our knowledge this is the first work suggesting such a link.} In Section \ref{whydeut} we outline why deuterium may matter for the IMF, and in Section \ref{wheredeut} show that deuterium abundance variations are expected to be strongest in the same locations where IMF variations are observed. We go on to discuss caveats, consequences, and future prospects in Section \ref{conclude}. Throughout the paper we use a $\Lambda$CDM cosmology with parameters $\Omega_\mathrm{m} = 1 - \Omega_\Lambda = 0.307$, $\Omega_\mathrm{b} = 0.04825$, $h = 0.6777$, $\sigma_8 = 0.8288$, $n = 0.9611$ \citep{Planck2014}, and assume the primordial abundance of deuterium with respect to hydrogen (D/H) is 25\,ppm \citep{2015PhRvD..92l3526C,2016RvMP...88a5004C}.

\section{Why Deuterium?}
\label{whydeut}

Deuterium is one of the few stable secondary isotopes produced in large amounts by big bang nucleosynthesis, together with $^{3}$He and $^{6}$Li (see \citealt{2007MNRAS.378..576S} for a review). While the abundance of helium and lithium can increase at later times (due to stellar nucleosynthesis and cosmic ray spallation) the abundance of deuterium can only decrease. 
As detailed below, protostars destroy all the deuterium they form with before they reach the main sequence, owing to the low temperature required for it to undergo fusion processes ($\approx10^6$ K).
Any deuterium produced in stellar nucleosynthesis is immediately destroyed for the same reason. 
Thus any gas which has been processed by and ejected from stars (i.e. mass lost in stellar winds or supernovae) is entirely deuterium-free, at least until it mixes with more pristine material.

As a protostar forms and accretes from its surrounding envelope it goes through five main stages (\citealt{1969MNRAS.145..271L,1980ApJ...236..201W,1991ApJ...375..288P}). Initially the protostar collapses under its own gravity, slowly increasing its core temperature (phase \textsc{I}). Once the core temperature reaches $\approx$1.5$\times10^6$\,K the star starts to burn deuterium (phase \textsc{II}). During the time taken to exhaust the deuterium in the core of the star, the central temperature is effectively held constant (by the extreme temperature sensitivity of this reaction which scales as $\sim T^{12}$; \citealt{1983ARA&A..21..165H}) while the star continues accreting mass. 

Once the deuterium in the central region of the protostar is exhausted, the star begins to contract further, allowing deuterium burning in shells at progressively larger radii (phase \textsc{III}). This release of nuclear energy in the lower density subsurface regions results in a swelling of the star. Once the entire body of the star has burned all available deuterium, it will again contract under gravity (phase \textsc{IV}), burning any deuterium it accretes almost instantaneously, until its core temperature reaches $\approx10^7$ K, when hydrogen fusion can begin. At this point, the star reaches the zero-age main sequence (ZAMS; phase \textsc{V}). Massive stars are known to continue accreting after the ZAMS is reached. However, the position of the `birth line' for lower mass protostars (where they become optically visible) suggests their birth cloud is typically dispersed (and thus they stop accreting from the environment) before they reach the ZAMS \citep[e.g.][]{2019A&A...624A.137H}. 
 
Given the key role of deuterium in the above process, it is clear that substantially reducing its abundance could strongly change the evolution of protostars. In the case where deuterium is absent they would skip all the intermediate stages (\textsc{II}-\textsc{IV}) described above and simply continue to contract. This substantially decreases the time available for the protostar to accrete from its surroundings before it joins the ZAMS.

In practise, the evolutionary sequence described above can be influenced by a variety of factors, including the specific entropy of the core from which the star forms, variations in the accretion flow, and the amount of energy input into the stellar atmosphere by accretion processes (see e.g. \citealt{2009ApJ...702L..27B, 2010ApJ...721..478H,Kunitomo2017}). This latter process is especially important if the accretion is (quasi-)spherical, because substantial amounts of accretion radiation will be trapped by the in-falling material. This additional source of energy can cause the star to swell and increases the time it takes to reach phase V (which could also affect the IMF). All of the above processes may vary, both within individual star-forming molecular clouds and on larger scales (e.g. for star formation taking place under different physical conditions). Exploring the full range of these variations, and the effect they may have on the IMF, is beyond the scope of this work. We here restrict ourself to a set of simple cases, which illustrate the effect deuterium abundance can have on the ZAMS mass of stars when other factors are kept constant. We return to the potential affect of the processes we do not study explicitly in Section \ref{discuss}.

We demonstrate the sensitivity of the zero-age main sequence mass of protostars to deuterium abundance by evolving accreting model protostars with the Modules for Experiments in Stellar Astrophysics \citep[\textsc{MESA}:][]{Paxton2011, Paxton2013, Paxton2015, Paxton2018, Paxton2019} code version 12778, following the procedure of \cite{Kunitomo2017}. The \textsc{MESA} code uses a combined equation-of-state from \citet{Rogers2002}, \cite{Saumon1995}, \cite{Pols1995}, \cite{Timmes2000} and \cite{Potekhin2010}. Radiative opacities are taken from \cite{Iglesias1993}, \cite{Iglesias1996}, \cite{Ferguson2005} and \cite{Buchler1976}.  Electron conduction opacities are from \cite{Cassisi2007}. Nuclear reaction rates are a combination of rates from \cite{Angulo1999}, \cite{Cyburt2010}, \cite{Fuller1985}, \cite{Oda1994} and \cite{Langanke2000}. Screening is included via the prescription of \cite{Chugunov2007}.  Thermal neutrino loss rates are from \cite{Itoh1996}.

\subsection{Deuterium variation at fixed initial core mass}

In our initial experiment, we explore the effect of the deuterium abundance in a regime where low-mass stars form from a core with the same initial mass and radius. We assume that accretion is steady and continues until the star reaches the ZAMS and arises from a disc with an inner cavity (where the majority of the accretion luminosity is radiated away into the surroundings). Because we are interested in studying the effect of deuterium, we do not attempt to model the initial stages of protostellar collapse from the molecular cloud or the formation of the second hydrostatic core. 
We instead begin our simulations when our protostars have reached a mass of 0.2\,\msun, using the \textit{create\_pre\_main\_sequence\_model} option in \textsc{MESA} \citep{1998ApJ...497..253U}.  These starting models have a uniform composition, a core temperature below $10^5$ K so no nuclear burning is occurring, and are uniformly contracting, producing enough luminosity to become fully convective. This choice of initial conditions can change the evolution of the stars somewhat (see Section \ref{dvar_smallcore} and e.g. \citealt{Kunitomo2017}). However, the resulting trends remain the same. Each protostar is set up so that all elements have solar abundance ratios (from \citealt{1998SSRv...85..161G}), apart from deuterium, whose abundance with respect to hydrogen is set to either 25,\,12.5,\ 2.5\ or\,0\,ppm.
We then evolve these models within \textsc{MESA} as they accrete from a disc at rates between $10^{-5}$ and $10^{-7}$\,\msunyr. We follow the evolution of each protostar until it reaches the ZAMS.

  \begin{figure}
\begin{center}
\includegraphics[width=0.48\textwidth,angle=0,clip,trim=0.0cm 0.4cm 0cm 0.0cm]{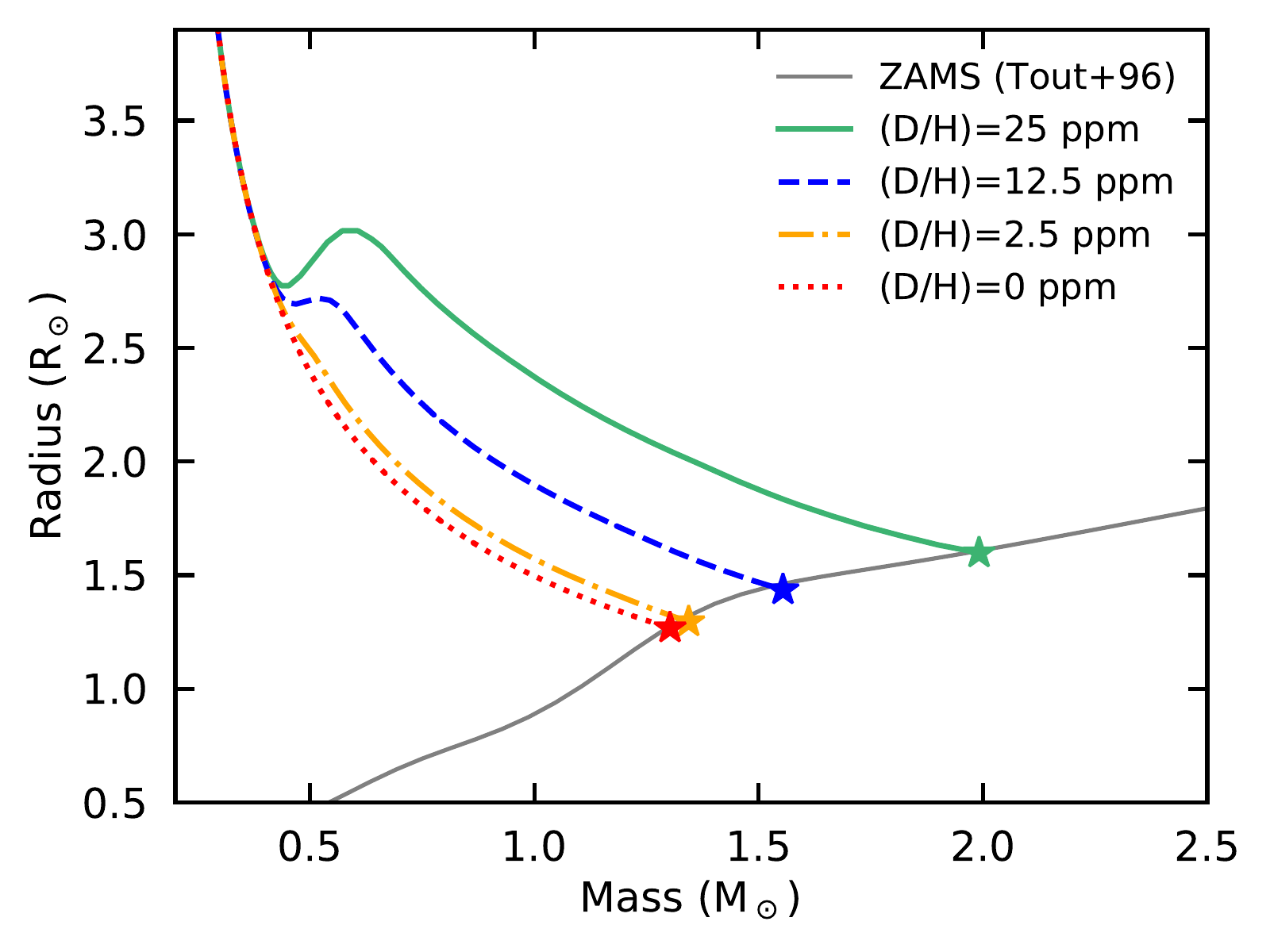}
\caption{Mass and radius evolution for a protostar of initial mass 0.2\,\msun\ accreting from a disc at $10^{-5}$\,\msunyr. Different curves show how this evolution changes if the protostar forms from material with different deuterium abundances (25 ppm; green solid curve, 12.5 ppm; blue dashed curve, 2.5 ppm; orange dot-dashed curve and deuterium free; red dotted curve). A star indicates where the protostar reaches the ZAMS (shown here as a grey solid curve; \protect \citealt{1996MNRAS.281..257T}), which happens at very different masses depending on the deuterium abundance.  }
\label{kunitomo}
 \end{center}
  \vspace{-0.6cm}
 \end{figure}

  \begin{table*}
\caption{Results from our MESA models of low-mass protostars evolving with different deuterium abundances.}
\begin{center}
\begin{tabular*}{0.8\textwidth}{@{\extracolsep{\fill}}l l r r r r r}
\hline
& M$_{\rm init}$ & R$_{\rm init}$ & log$_{10} \dot{\mathrm{M}}$ & [D/H] & M$_{\rm ZAMS}$ & $\Delta$M\\
 & (M$_{\odot}$) & (R$_{\odot}$) & ( M$_{\odot}$ yr$^{-1}$) & (ppm) & (M$_{\odot}$) & (M$_{\odot}$)\\
& (1) & (2) & (3) & (4) & (5) & (6) \\
\hline
\multicolumn{6}{l}{1) Accretion variation at fixed initial mass:} \\
& 0.2 &  23.2 & -5 & 25.0 & 1.95 & -- \\
& 0.2 &  23.2 & -5 & 12.5 & 1.52 & 0.43 \\
& 0.2 &  23.2 & -5 & 2.5 & 1.32 & 0.63 \\
& 0.2 &  23.2 & -5 & 0.0 & 1.27 & 0.67 \\
& 0.2 &  23.2 & -6 & 25.0 & 1.11 & -- \\
& 0.2 &  23.2 & -6 & 12.5 & 1.02 & 0.09 \\
& 0.2 &  23.2 & -6 & 2.5 & 0.94 & 0.17 \\
& 0.2 &  23.2 & -6 & 0.0 & 0.90 & 0.20 \\
& 0.2 &  23.2 & -7 & 25.0 & 0.74 & -- \\
& 0.2 &  23.2 & -7 & 12.5 & 0.69 & 0.05 \\
& 0.2 &  23.2 & -7 & 2.5 & 0.64 & 0.10 \\
& 0.2 &  23.2 & -7 & 0.0 & 0.63 & 0.11 \\
\hline
\multicolumn{6}{l}{2) Smaller initial mass:} \\
& 0.01 & 1.5 & -5 & 20.0 & 0.84 & --\\
& 0.01 & 1.5 &-5 & 10.0 & 0.60 & 0.24\\
& 0.01 & 1.5 & -5 & 0.0 & 0.45 & 0.39\\
\hline
\multicolumn{6}{l}{3) Time varying accretion rate:} \\
& 0.2 & 23.2 & V\&B15 & 25 & 0.84 & --\\
& 0.2 & 23.2 &V\&B15 & 12.5 & 0.67 & 0.17\\
& 0.2 & 23.2 &V\&B15 & 2.5 & 0.62 & 0.22\\
& 0.2 & 23.2 & V\&B15 & 0 & 0.61 & 0.23\\
\hline
\end{tabular*}
\parbox[t]{0.8\textwidth}{ \textit{Notes:} Column one and two show the initial mass and radius of the core in place at the start of the protostellar model run. Model types 1 and 3 have core masses and radii following the scaling relations of \protect \cite{1998ApJ...497..253U}, while model type 2 uses the initial conditions from \cite{Kunitomo2017}. Column 3 shows the accretion rate onto the core, which is constant for model types 1 and 2, but varies in model 3 following the prescriptions of \protect \cite{2015ApJ...805..115V}. Column 4 shows the initial deuterium abundance of each model. Column 5 shows the mass of each star as it reached the ZAMS, while Column 6 shows the difference between that mass and the mass of the equivalent star with a primordial deuterium abundance. Note that for model type 2 we calculate these values with respect to the 20 ppm deuterium (instead of 25 ppm) star due to the different initial deuterium abundance values used in \cite{Kunitomo2017}. In all cases, stars formed from lower deuterium abundance material have lower ZAMS masses and this effect is more pronounced in higher mass stars.}
\end{center}
\label{deut_table}
\end{table*}

The results of these model runs are tabulated in Table \ref{deut_table}, while in Figure \ref{kunitomo} we show the result of these calculations for objects with a range of deuterium abundances accreting at $10^{-5}$\,\msunyr (which we call our reference model in what follows). 
The mass and radius evolution of these objects are shown as different curves in Figure \ref{kunitomo}, which are coloured by their deuterium abundance from 25\,ppm (solid green curve), 12.5\,ppm (dashed blue curve), 2.5\,ppm (dot-dashed orange curve) and zero\,ppm (dotted red curve). 

In this reference model deuterium burning begins after $\approx$10,000 years, and results in a characteristic swelling of the star. Deuterium in each star is exhausted fairly quickly, at which point it begins to contract towards the ZAMS again. A coloured star indicates the point at which hydrogen burning begins in the core of each star, which well reproduces the ZAMS mass-radius relation for this metallicity from \cite{1996MNRAS.281..257T}, shown as a solid grey curve.  

It is clear from Figure \ref{kunitomo} that different deuterium abundances result in wildly different ZAMS masses; the star with a primordial (D/H)=25\,ppm has a mass of $\approx$\,2.0\,\msun, while the deuterium-free star has a mass of only $\approx$\,1.3\, \msun. The same trends are shown in Table \ref{deut_table} for other accretion rates. In each case the lower the deuterium abundance, the lower the ZAMS mass of the resultant star. Interestingly the fractional difference between the mass of the star with primordial deuterium abundance and stars with lower abundance appears to increase with mass. Although these models are fairly simplistic, this suggests that the slope of IMF would be altered by a change in deuterium abundance, and become more bottom heavy, as observed in massive ETGs. 

While disc accretion can continue after the ZAMS, observations and simulations suggest it is difficult for low-mass main-sequence objects to accrete significant amounts of material, as it must overcome the additional radiation and stellar-wind pressure generated by hydrogen fusion (see e.g. \citealt{2016ARA&A..54..135H}). It thus seems unlikely that such large differences in ZAMS mass can be systematically erased by future accretion

\subsection{Initial core mass variations}
\label{dvar_smallcore}
As discussed above, the specific entropy of the core from which a protostar forms has an impact on its ZAMS mass. We thus repeat the calculation performed above, but using lower mass seeds (0.01 \msun; 10.5 M$_{\rm Jup}$) taken from the work of \cite{Kunitomo2017}. These seeds accrete from a disc at the same rate as the reference model described above ($10^{-5}$ \msunyr), and have initial deuterium abundances of 20\,ppm, 10\,ppm, and zero\,ppm. The results of these calculations are again tabulated in Table \ref{deut_table}. Although the ZAMS mass of the star is very different from our reference calculation in this scenario (as expected given the higher density of the initial seed) stars formed from deuterium-depleted material again have significantly lower masses. 

\subsection{Time varying accretion rates}
 
In the above sections we have assumed that stars accrete steadily from their birth cloud. In reality temporal variations in accretion rate are significant, and it is likely that the majority of the mass is accreted onto low-mass protostars early in their evolution. We test the impact of this by running a series of models as above, but using accretion histories parameterised from the disk gravitational instability and fragmentation models of \cite{2015ApJ...805..115V}. In these models accretion begins at a rate $\approx10^{-5}$ \msunyr, and declines as $\approx t^{-1/3}$.

We implemented this accretion evolution in MESA, and evolved a set of protostars (whose other parameters such as initial mass and radius were kept fixed to the values used in our reference model). The results of these calculations are  tabulated in Table \ref{deut_table}. Once again the lower deuterium abundance protostars have significantly lower ZAMS masses than those formed from primordial material, suggesting accretion rate variations alone cannot remove the effect. 

We note that the time taken for a given mass protostar to reach the deuterium burning phase is dependent on its total mass (and thus its accretion rate for any given initial core mass). Higher mass systems evolve more quickly, and thus the presence of a deuterium burning phase has a maximal chance to affect the evolution of high-mass stars. Lower mass protostars evolve more slowly, and thus may have accreted the majority of their mass before deuterium burning begins. This once again suggests that deuterium abundance changes are likely to reduce the masses of higher mass protostars more than lower mass systems, and thus potentially create a bottom-heavy IMF.

\section{Where do deuterium variations matter?}
\label{wheredeut}

We have shown above that deuterium abundance variations have the potential to change the ZAMS mass of accreting protostars such that the IMF may become more bottom-heavy. In this Section we show that deuterium variations are expected to be important in the same settings where a bottom-heavy IMF has been observed - in the centres of massive passive galaxies.

Unfortunately the amount of deuterium present in forming protostars leaves little, if any, observable trace in main-sequence stellar properties. Compounding this issue, the deuterium abundance in the ISM itself is hard to measure in other galaxies (with the exception of some quasar sightline systems). As such we here turn to cosmological hydrodynamic simulations. 

\subsection{The EAGLE simulations}

  \begin{figure*}
\begin{center}
\includegraphics[width=1\textwidth,angle=0,clip,trim=0.0cm 0.4cm 0cm 0.0cm]{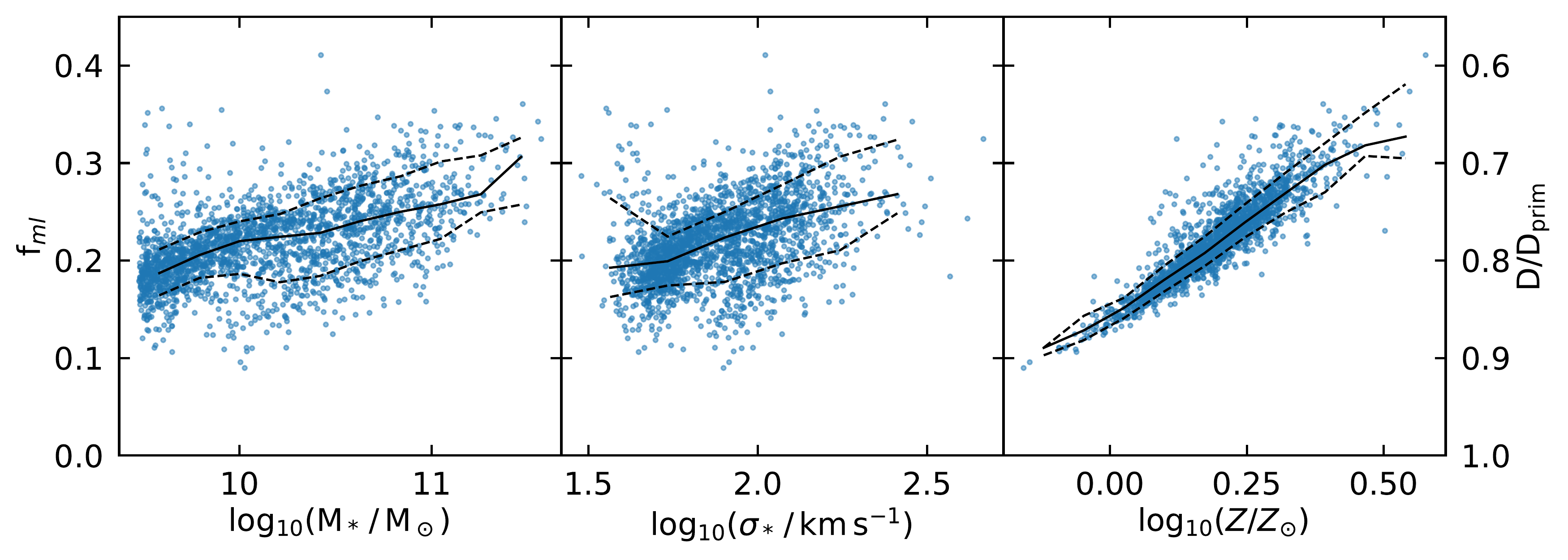}
\caption{Median fraction of deuterium-free mass loss incorporated in stars that lie within the central 10 kpc of passive galaxies (defined as having a SFR/M$_{*}<10^{-11}$\,yr$^{-1}$) with stellar masses $>3\times10^9$\msun\ at $z=0$ in EAGLE. These are plotted against the total stellar mass, velocity dispersion, and metallicity of these systems (blue points). A second $y$-axis shows the median deuterium fraction (with respect to primordial) at the birth of these stars. Solid and dashed black curves indicate the median and 16th and 84th percentiles of the distribution in each panel. High mass, high velocity dispersion, and high metallicity galaxies (the types of systems where IMF variations have been observed) have a higher median mass loss fraction and thus a lower average deuterium abundance.}
\label{deut_galprops_eagle}
 \end{center}
   \vspace{-0.6cm}
 \end{figure*}

We make use of the `Evolution and Assembly of GaLaxies and their Environments' (EAGLE) project \citep{Schaye2015, Crain2015, 2016A&C....15...72M}, which consists of a large number of cosmological. hydrodynamical simulations with different subgrid physics, simulation volumes, and resolutions, which were run using a modified version of the \textsc{gadget-3} code \citep[last described in][]{Springel2005}. Here we make use of the largest (100 co-moving Mpc on a side) cubic periodic volume simulation (\emph{``Ref-L100N1504''}), which includes the effect of both stellar and active galactic nucleus (AGN) feedback. A comprehensive description of this simulation, and of the database of galaxy properties extracted from it, can be found in \cite{Schaye2015}, \cite{Crain2015}, \cite{2016A&C....15...72M}, and \cite{2017arXiv170609899T}.

This simulation uses a smoothed particle hydrodynamics (SPH) scheme to follow the gaseous components of the Universe. This has significant advantages over grid codes for our application (as well as some disadvantages). In EAGLE, each gas particle begins with a fixed initial mass ($M_{\rm min}$ = 1.81$\times10^6$\,\msun\ in the reference simulation used here), which can increase as stellar mass loss material is incorporated. with that particle (there is no mixing). 
Each star particle that forms represents a stellar population of a single age (a simple stellar population; SSP) which inherits its mass and metallicity from its progenitor gas particle. When a star forms from a gas particle it thus encodes the fraction of mass loss material present in the gas ($f_{\rm ml}$), which directly relates to the deuterium fraction of that gas with respect to primordial ($\frac{\mathrm{D}}{\mathrm{D_{prim}}}$).
This deuterium fraction is encoded in the ratio of the initial mass of the star particle ($M_{\rm *, init}$) to the minimum particle mass ($M_{\rm min}$):
\begin{eqnarray}
f_{\rm ml} = 1 - \frac{M_{\rm min}}{M_{\rm *, init}} = 1-\frac{\mathrm{D}}{\mathrm{D_{prim}}}. 
\end{eqnarray}

The method described above has been used to study the evolution and importance of mass loss in EAGLE \citep{2016MNRAS.456.1235S}.  A similar scheme has been used to study deuterium in cosmological zoom simulations \citep{2018MNRAS.477...80V}. 

EAGLE adopts a \cite{2003PASP..115..763C} IMF, spanning the mass range of 0.1 -- 100 M$_{\odot}$. Following the prescriptions of \cite{2009MNRAS.399..574W}, a star particle loses mass through stellar winds, core collapse and Type Ia supernova explosions. The rate at which mass is lost is calculated using the metallicity-dependent stellar lifetime tables of \cite{1998A&A...334..505P}. As this mass is lost from the stellar particles, it is allocated to the nearest 48 gas particles according to the smooth particle hydrodynamics interpolation scheme. This interpolation scheme has a disadvantage for our study, because there is no limit on the distance to those gas particles. In a gas-rich galaxy this is unlikely to matter. However in a quenched early-type system with little gas, this mass loss may end up being added to hot halo particles far from the galaxy centre, rather than forming a new reservoir of gas within the galaxy. This likely causes us to underestimate the true contribution of mass loss to the ISM (and thus star formation from deuterium-depleted material) in some circumstances.

In Figure \ref{deut_galprops_eagle} we show the median mass loss fraction of the stars that lie within the central 10 kpc of passive galaxies (defined as having a SFR/M$_{*}<10^{-11}$\,yr$^{-1}$) with stellar masses $>3\times10^9$ \msun\ at $z=0$ in EAGLE. This median mass loss fraction corresponds directly to the deuterium fraction (with respect to the primordial) of these stars at the moment they were formed (shown as a second y-axis). We plot these quantities as a function of the total stellar mass, velocity dispersion, and metallicity of these systems (blue points). The median, 16th, and 84th percentiles of the distribution are shown as black solid and dashed curves, respectively. 
 The most massive, highest velocity dispersion, and highest metallicity objects show the highest mass loss fraction. They thus would have formed stars with the lowest average deuterium content (in agreement with the results of \citealt{2016MNRAS.456.1235S} and \citealt{2018MNRAS.477...80V}).

   \begin{figure}
\begin{center}
\includegraphics[width=0.5\textwidth,angle=0,clip,trim=0.0cm 0.4cm 0cm 0.0cm]{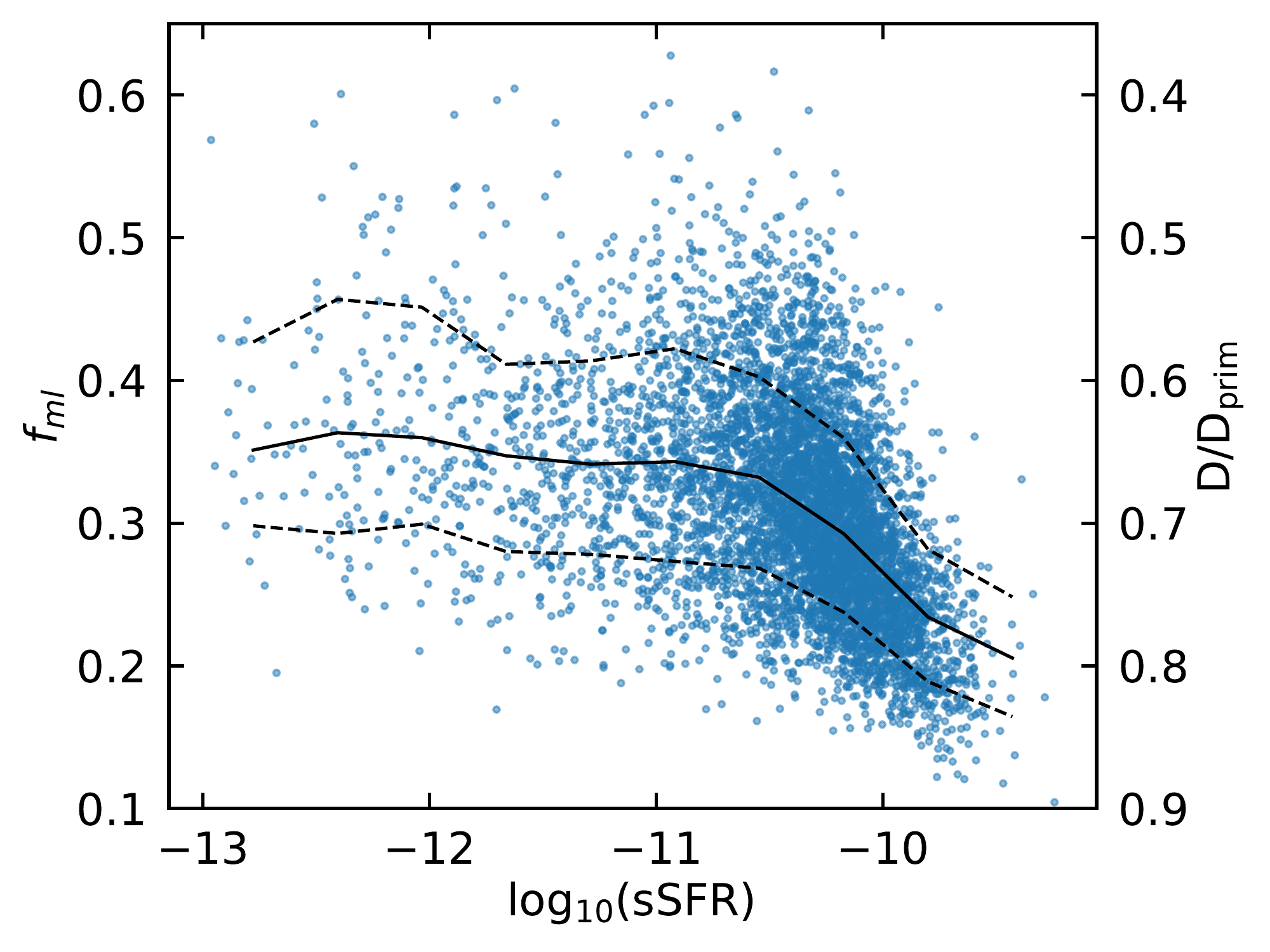}
\caption{{The median fraction of deuterium-free mass loss incorporated into stars that have formed since $z=1$ in each galaxy extracted from EAGLE, plotted against its $z=0$ specific star formation rate (blue points). Solid and dashed black curves indicate the median and 16th and 84th percentiles of the distribution. Objects that are strongly star-forming today have had significantly more deuterium in their ISM over the last $\approx$8 Gyrs, while $z=0$ passive systems have been forming stars from material with significantly lower deuterium abundance.}}
\label{deut_ssfr_eagle}
 \end{center}
   \vspace{-0.6cm}
 \end{figure}

{In Figure \ref{deut_galprops_eagle} we only considered passive galaxies. In order to investigate what happens in more star-forming systems in Figure \ref{deut_ssfr_eagle} we show the median fraction of deuterium-free mass loss in stars that have formed since $z=1$ in each galaxy, plotted against their $z=0$ specific star formation rate (blue points). The objects that have formed a large fraction of their stellar mass recently have done so from much more deuterium-rich gas than those that are more passive. If the deuterium abundance is an important parameter in mediating IMF variation then this may help explain why altered IMFs are primarily observed in passive systems. This is discussed further in Section \ref{discuss}.} 

\begin{figure}
\begin{center}
\includegraphics[width=0.5\textwidth,angle=0,clip,trim=0.0cm 0.4cm 0cm 0.0cm]{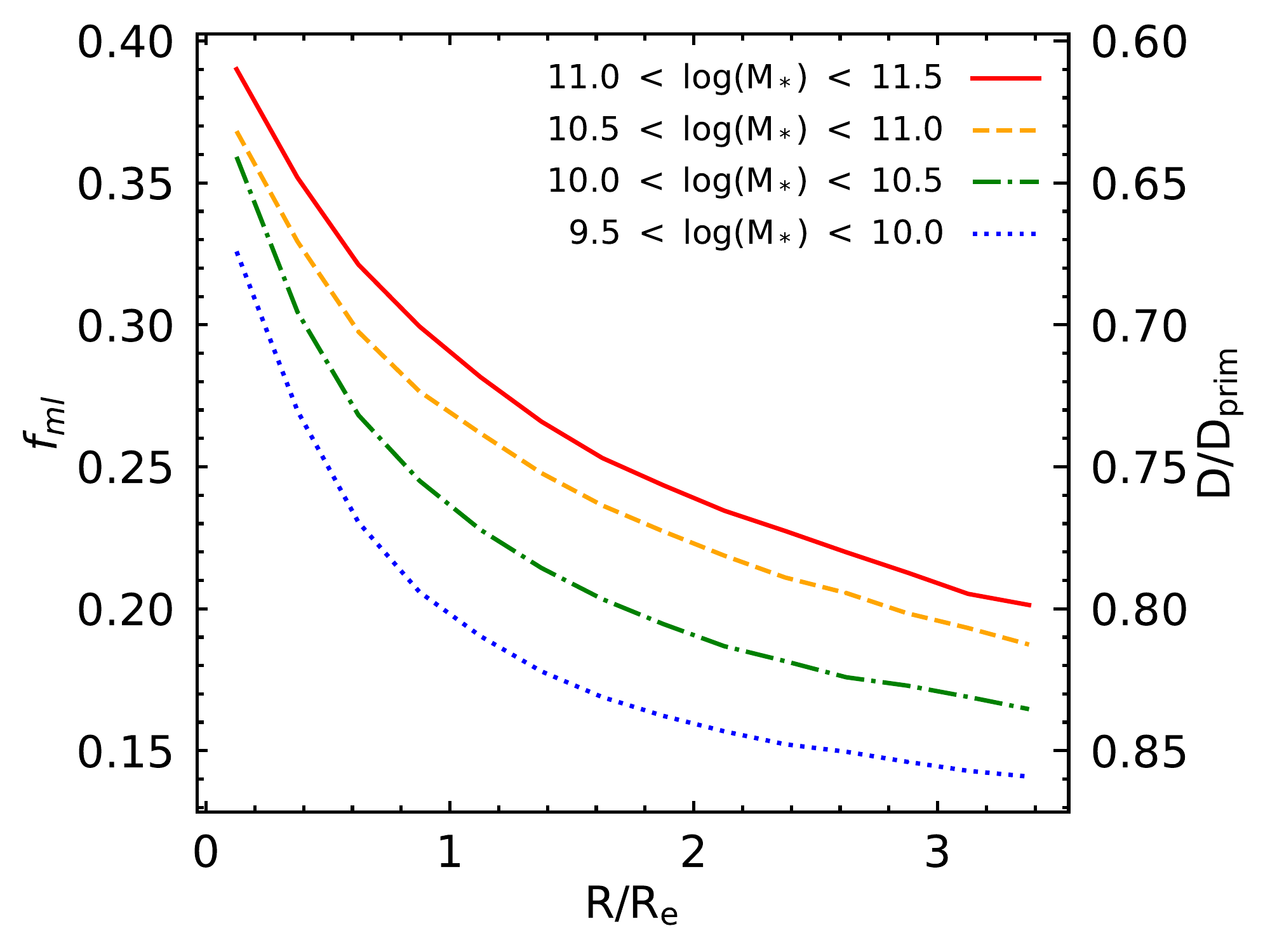}
\caption{Median fraction of deuterium-free mass loss incorporated in stars as a function of radius {(normalised by the half-mass radius, here denoted R$_{\rm e}$)} within passive galaxies in EAGLE.  Objects with $9.5<\log(\mathrm{M}_*/\mathrm{M}_\odot)<10.0$ are shown as a dotted blue curve, while objects with $10.0<\log(\mathrm{M}_*/\mathrm{M}_\odot)<10.5$ are shown as a dash-dotted green curve, $10.5<\log(\mathrm{M}_*/\mathrm{M}_\odot)<11.0$ as a dashed orange curve, and $11.0<\log(\mathrm{M}_*/\mathrm{M}_\odot)<11.5$ as a solid red curve. Deuterium-free mass loss is important in the central regions of passive galaxies at all masses.}
\label{deut_radial_eagle}
 \end{center}
   \vspace{-0.6cm}
 \end{figure}

In Figure \ref{deut_radial_eagle} we show how the stars created from deuterium-poor mass loss material are distributed in their passive parent galaxies. We plot the median mass loss fraction as a function of radius {(normalised by the half-mass radius)} for objects in different stellar mass bins (as indicated in the legend). In all cases the mass loss fraction is highest in the central regions (a median mass loss fraction of between 33-41 per cent depending on stellar mass) and decreases radially. Passive galaxies with different stellar masses show similar deuterium gradients, but with a different normalisation (as expected from Figure \ref{deut_galprops_eagle}).  

\begin{figure}
\begin{center}
\includegraphics[width=0.48\textwidth,angle=0,clip,trim=0.0cm 0.5cm 0cm 0.0cm]{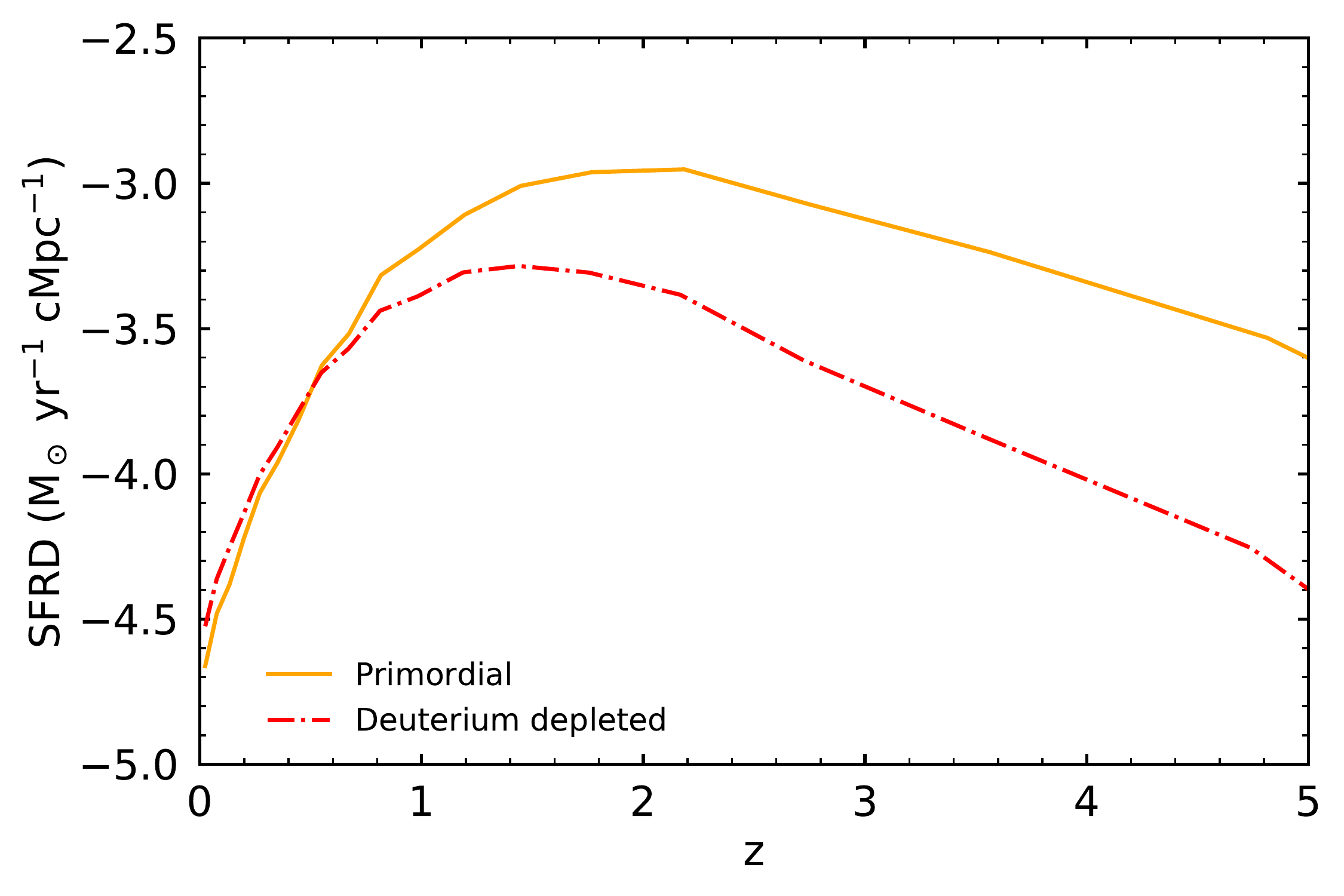}
\caption{The evolution of the cosmic SFR density for stars which lie in the inner kiloparsec of massive ($>10^{10}$\,\msun), passive (SFR/M$_{*}<10^{-11}$\,yr$^{-1}$) galaxies at redshift zero in EAGLE. Curves are shown for star formation fuelled by gas that has, and has not, been recycled through stars (red dashed curve and orange solid curve, respectively). The star formation rate density fuelled by processed deuterium-free material is highest at $z\approx1.5$ and begins to dominate the total star formation rate density of such massive, passive systems at $z<0.5$.  }
\label{deut_sfh_eagle}
 \end{center}
 \vspace{-0.7cm}
 \end{figure}

Another question the simulations allow us to address is the cosmic evolution of (deuterium-free) stellar mass loss fuelling star formation. 
\cite{2016MNRAS.456.1235S} showed that the fractional contributions of stellar mass loss to the cosmic star formation rate increases with time, reaching 35 per cent at $z = 0$ when averaged over the entire galaxy population. In Figure \ref{deut_sfh_eagle} we explore this further, extracting from the EAGLE simulation the star formation history of stars that lie in the central kiloparsec of massive galaxies ($>10^{10}$\,\msun) which are passive (SFR/M$_{*}<10^{-11}$\,yr$^{-1}$) at the present day.  The fraction of the star formation rate density which arises from primordial material is shown as a solid orange curve, while the fraction of star formation that is fuelled from processed (deuterium-depleted) material in these systems is shown as a dash-dotted red curve. The total amount of star formation arising from deuterium-depleted stellar mass loss peaks at $z\approx1.5$ in these passive systems (where it provides $\approx$30 per cent of the total star formation rate density; SFRD). In addition, mass loss dominates the total star formation rate density in the central kiloparsec of these galaxies at $z\ltsimeq0.5$ (contributing $>$60 per cent of the SFRD by $z=0$).

Overall, analysis of these cosmological simulations show us that deuterium variations are expected in the same regimes where IMF variations have been found. The strongest deuterium depletion is seen in the centres of massive, metal enriched galaxies, which is where IMF variations are observed. Deuterium-poor gas is available in significant quantities even 10 Gyr ago (when the bulk of the stars in ETGs are forming) and becomes even more important in the centres of massive galaxies at low redshifts.  We will discuss this, and its implications, further in Section \ref{conclude}.

\section{Discussion and Conclusions}
\label{discuss}
\label{conclude}

Understanding the cause of any variation in the IMF is clearly of great importance if we want to constrain galaxy formation theories with electromagnetic observations. In this paper we have shown that the deuterium abundance is one physical parameter that could have an effect on the IMF. 

We have used MESA models of protostars to show that those forming from deuterium-free material would be expected to {be} significantly less massive than those forming under the same conditions but with higher deuterium abundances. The sensitivity of protostellar evolution to deuterium abundance changes in stars of different masses in our model, in a way that would likely create a bottom-heavy IMF.

Various uncertainties could impact this result. For instance, the specific entropy of the initial pre-main sequence seed can change the evolution of the protostars somewhat (as explored in Section \ref{dvar_smallcore}), although for any initial specific entropy the deuterium abundance effect remains. The specific entropy distribution of pre-stellar cores may well change in the highly turbulent gas of starburst galaxies, but it seems unlikely that it would vary in the exact way required to systematically remove any effect from the deuterium abundance. 

A further uncertainty arises from our treatment of accretion. Here we have studied the cases where accretion is steady state or declining with time, and arises from a disc with an inner cavity that lets the majority of the accretion luminosity escape. If a large amount of accretion energy were to instead be trapped, then it can swell the star and reduce the dependence of the the IMF on the deuterium abundance (see e.g. \citealt{2009ApJ...691..823H}). 
In star-forming regions within the Milky Way, it seems that changes in the deuterium abundance alongside changes in accretion energy trapping may help to explain the full spread of protostar luminosities observed \citep{Kunitomo2017,2018MNRAS.474.1176J}. 
However, it is unclear under which conditions each of these effects dominate. Some studies find that accretion energy trapping may be more common at higher accretion rates \citep{2018MNRAS.474.1176J}. Thus this accretion energy trapping effect could also potentially affect the IMF. Further exploration of the interplay between these processes, and how they manifest in real star forming environments is clearly required to fully understand this.

Finally, our simple models are only valid for low-mass stars ($\ltsimeq2$\,\msun). Forming higher mass stars following these prescriptions is not possible without inputing exceedingly high accretion rates (significantly higher than those expected observationally and theoretically; e.g. \citealt{2014ApJ...797...32P}). Massive star formation is known to proceed differently from low mass star formation \citep[e.g.][]{2016IAUS..315..154T,2018ARA&A..56...41M}, and may be strongly influenced by the local environment \citep[e.g.][]{2013A&A...555A.112P,2018A&A...613A..11W,2018A&A...610A..77H}. We thus do not speculate here on the effect deuterium abundance could have on these systems. Even if these massive stars were immune to any effect, reducing the amount of $\approx$0.5-2 \msun\ stars (creating a `scooped' IMF) could still affect observations, as these stars dominate the mass, and in passive galaxies are typically the most massive stars still present (and thus dominate their spectra). 

{In Section \ref{wheredeut} we used a simulation from the EAGLE suite to explore where variations in the deuterium fraction are important. We showed that the deuterium abundance is expected to be lowest in massive metal-enriched passive galaxies. We also showed that there is significant radial variation in the deuterium fraction within galaxies, with their centres showing the largest deuterium depletion. Thus we predict that deuterium abundance variations occur exactly where evidence for IMF variation is strongest.} 

{The radial deuterium abundance gradients we find in the simulated EAGLE galaxies are, however, very similar for passive galaxies of all stellar masses, just with a different normalisation that depends on mass. This is somewhat at odds with observations, which typically find shallow or no radial gradients in the IMF of low mass ETGs \citep[see e.g.][]{2018MNRAS.477.3954P}. Observational studies which trace radial variation in the IMF are in their infancy and suffer from significant projection effects. Assuming the observed lack of IMF gradients in low-mass galaxies is correct, however, then (if deuterium is linked to IMF variation) perhaps the deuterium abundances found in these lower mass galaxies are high enough even at their centres to avoid significant IMF variation. Alternatively, it is possible that the limitations of the simulation may be causing this effect. For instance, no metal mixing occurs between individual resolution elements in the simulations. If local mixing were included it would likely reduce radial abundance variations. This effect is likely to be strongest in the low mass systems, which are physically smaller and thus mix more easily. Including the effect of local mixing while tracing deuterium explicitly in future simulations would help constrain this.}

Many authors have assumed that the starbursts that create massive ETGs are where the process(es) that change the IMF manifest themselves. This is supported by the correlation reported by some authors between the IMF and alpha-element abundance \citep[e.g.][]{2012ApJ...760...71C}. High alpha-element abundances suggest that the bulk of the stars in these systems formed quickly, before supernova type-Ia's pollute the ISM with significant amounts of iron. 
Thus it has been assumed that objects with the most bottom-heavy IMFs formed the majority of their mass very quickly.
If deuterium were the cause of IMF variation, however, it seems more likely that the stars formed initially in the starburst would have a more normal IMF, and the stars that are formed later may be the ones with a different IMF. 
These deuterium-poor stars could have been formed right after the initial burst of star formation when the alpha-enrichment would have still been high (as $\approx$25\% of the initial mass of stars returns to the ISM as deuterium free gas within 1 Myr of the onset of a star formation burst, and $\approx$40\% within a Gyr; see e.g. Fig.~1 in \citealt{2016MNRAS.456.1235S}). Alternatively they could have been added at later times as stellar mass loss from lower mass AGB stars accumulated. This latter processes is expected to be important in fuelling the star formation observed in massive galaxies (see e.g. Section \ref{wheredeut}; \citealt{2016MNRAS.457..272D,2019MNRAS.486.1404D}). 

{We note that different IMFs have different mass-loss rates. For instance, a \cite{2003PASP..115..763C} IMF (as assumed in EAGLE) returns around 1.5 times more mass to the ISM than a \cite{1955ApJ...121..161S} IMF (see e.g. Appendix A of \citealt{2016MNRAS.456.1235S} and references therein). Assuming deuterium (and thus stellar mass-loss material) is important in altering the IMF, this means that mass-loss rates will slow somewhat as successive generations of stars form with more bottom-heavy (Salpeter-like) IMFs. 
The impact of this is hard to gauge and will depend on the channel which forms the majority of the IMF-discrepant stars in present-day galaxies. For instance, the lower mass-loss rate of stars with a \cite{1955ApJ...121..161S} type IMF would not impact the eventual composite IMF of a galaxy if the majority of the stars with a bottom-heavy IMF formed preferentially in a single burst (e.g. as a second-generation population within a starburst). If the more secular channels of IMF change discussed above dominate, however, then this could slow the accumulation of new deuterium-free gas, and thus the production of new stars with an altered IMF. A better understanding of this should be possible within simulations which explicitly trace the changing IMF and its effect on mass loss and deuterium enrichment. }

If the IMF of forming stars were to vary temporally as described above then this has several implications for our understanding of galaxy evolution. 
For instance, the colours and optical spectra of ETGs suggest (when analysed assuming a fixed IMF) that little star formation has occurred in these systems in the last $\approx$10\,Gyr \citep[e.g.][]{2005ApJ...621..673T}. This is despite galaxy mergers, stellar mass loss, and accretion providing ample material from which stars could form. This has been used to argue for efficient `maintenance mode' AGN feedback which heats and removes gas from the ISM and halo of massive galaxies \citep[e.g.][]{2006MNRAS.370..645B,2006MNRAS.365...11C}. If the IMF were to vary as described above, however, star formation fuelled by deuterium-depleted material at intermediate redshifts would be difficult to detect observationally in $z=0$ galaxies (as it would have a deficit of solar-type stars which would be expected to dominate present-day spectra). This may reduce the amount of maintenance mode AGN feedback required to match the properties of the observed galaxy population.

Additionally, star formation in local ETGs appears to proceed differently from star formation in local spiral galaxies, turning gas into stars with a significantly lower efficiency \citep{2011MNRAS.415...61S,2014MNRAS.444.3427D}. If this star formation were to be fuelled by deuterium-free stellar mass loss, and if that were to result in a significantly different IMF, this would bias the observations. It is then possible that there is no need for large scale star formation efficiency changes in this population. 

Although we are unable to determine the shape of the IMF that results from deuterium-depleted star-formation in this work, we do find hints that it could have a `scooped' shape, with massive stars being unaffected.
{Typically, stellar population analyses fit IMFs with single power-law shapes, or multiple power laws with free slopes between e.g. 0.1 - 0.5, 0.5 - 1 and 1-100 \msun. Our analyses suggest that deuterium abundance variations may primarily affect stars from 0.5 - 2 solar masses (although the exact mass range affected is assumption dependent, as discussed above). Some authors have suggested variation of the IMF within the 0.5 - 1 \msun\ region could provide the best fit to their observational data \citep{2020arXiv200809557F}, but further analysis is required to confirm this. Overall our results suggest that more complicated `scooped' shapes may need to be included in IMF studies in the future.} 

It is important to note that a `scooped' IMF shape, as discussed above, would cause dynamical measurements of the IMF mismatch parameter (which rely on mass-to-light ratios; see e.g. \citealt{2012Natur.484..485C}) to vary temporally. The inferred IMF mismatch parameter would depend both on the mass-fraction of stars with an altered IMF, and the time since those stars were formed. Some hints of the IMF mismatch parameter having a secondary age dependence have been found observationally in the past \citep[e.g.][]{2019MNRAS.485.5256Z}. This would be expected to cause significant scatter in IMF -- host galaxy correlations.

Deuterium-free stellar mass loss is an important source of fuel for star formation in galaxies of all types \citep[e.g.][]{2016MNRAS.456.1235S}. If deuterium were important in setting the IMF then one could question why the Milky Way and other spiral galaxies do not show differences in their IMF. {We showed (Figure \ref{deut_ssfr_eagle}) that high sSFR galaxies are more deuterium-rich material than passive galaxies.} We here posit that this is because of the smoother, accretion-dominated star formation histories of spiral galaxies. Such systems are thought to form stars at a fairly steady rate across cosmic time, mainly fuelled by accretion of metal-poor (and thus deuterium-rich) material \citep[e.g.][]{2012MNRAS.423.2991V}. The Milky Way appears to have followed this evolutionary path, with a current day gas accretion rate which is similar to its total star formation rate \citep[e.g.][]{2014ApJ...787..147F}. As star formation efficiencies in spiral galaxies are low (depletion times of $\approx$2 Gyr; e.g. \citealt{2011ApJ...730L..13B}) and mixing times are short \citep[$\approx$100 -- 350 Myr;][]{2002ApJ...581.1047D,2019ApJ...887...80K} deuterium-poor mass loss material likely has time to mix with newly accreted deuterium-rich gas before it forms stars. This is supported by the approximately constant (D/H) ratio measured across the Milky Way disc (once dust depletion is accounted for: e.g. \citealt{2006ApJ...647.1106L}). In starbursts at high redshift the depletion time can be much smaller than this mixing time and the return rate of deuterium-depleted material much higher than in a steady state system, perhaps allowing deuterium-depleted stars to form. ETGs at low redshift, on the other hand, have low gas return rates, but inflow rates that are even lower, potentially allowing them to host star formation from an entirely deuterium-depleted ISM. 

Additional work is clearly required to determine if deuterium does play a role in IMF variation. 
Given the difficulties in determining the deuterium fraction of gas in galaxies (especially given the selective fractionation effects that can affect cold ISM tracers) it is crucial to concentrate observationally on systems where we strongly suspect stellar mass loss is dominating the supply of gas. One such location is in satellite ETGs within galaxy clusters, where essentially all sources of external primordial gas have been removed. If late-time star formation with an altered IMF is important then these systems may show stronger IMF variation than their field counterparts, and with less scatter. A recent study of the Coma cluster \citep{2020MNRAS.494.5619S} finds a very tight IMF -- $\sigma_*$ correlation, that is slightly offset from relations primarily based on field objects. While these changes may also be due to systematic differences in analysis, it at least suggests this is a promising line of enquiry for future work. By combining such work with further advances in simulation techniques and more accurate protostellar modelling, the role (if any) of deuterium in IMF variation can be revealed.

\section*{Acknowledgements}

We thank the referee for helpful comments which improved this work.
TAD acknowledges support from the UK Science and Technology Facilities Council through grant ST/S00033X/1, and thanks Dr Masanobu Kunitomo for sharing his MESA routines and protostellar seeds. FvdV acknowledges support from a Royal Society University Research Fellowship. We thank Dr. Paul Clarke, Prof. Jane Greaves, Prof. Ant Whitworth, Dr. Nick Wright, and the stars group at Keele University for helpful discussions. 

\section*{Data avalability}
The data and scripts underlying this article are available via GitHub, at \url{https://github.com/TimothyADavis/DeuteriumIMF}.

\bsp	
\bibliographystyle{mnras}
\bibliography{bibimf_d.bib}
\bibdata{bibimf_d.bib}
\bibstyle{mnras}

\label{lastpage}
\end{document}